\documentclass[letterpaper]{article} % DO NOT CHANGE THIS
\usepackage[]{aaai23}  % DO NOT CHANGE THIS
\usepackage{times}  % DO NOT CHANGE THIS
\usepackage{helvet}  % DO NOT CHANGE THIS
\usepackage{courier}  % DO NOT CHANGE THIS
\usepackage[hyphens]{url}  % DO NOT CHANGE THIS
\usepackage{graphicx} % DO NOT CHANGE THIS
\usepackage{natbib}  % DO NOT CHANGE THIS AND DO NOT ADD ANY OPTIONS TO IT
\usepackage{caption} % DO NOT CHANGE THIS AND DO NOT ADD ANY OPTIONS TO IT
\usepackage{algorithm}
\usepackage{algorithmic}

\usepackage{newfloat}
\usepackage{listings}
\DeclareCaptionStyle{ruled}{labelfont=normalfont,labelsep=colon,strut=off} % DO NOT CHANGE THIS
\floatstyle{ruled}
\newfloat{listing}{tb}{lst}{}
\floatname{listing}{Listing}
\usepackage{multirow}
\usepackage{dblfloatfix}
\usepackage{algorithm}
\usepackage{algorithmic}
\usepackage{amsfonts}

\usepackage{newfloat}
\usepackage{listings}
\usepackage{color,soul}
\usepackage{multirow}

\begin{document}

\title{A Dataset and Baseline Approach for Identifying Usage States from Non-Intrusive Power Sensing With  MiDAS IoT-based Sensors}

\author{
    Bharath Muppasani \textsuperscript{\rm 1},
    Cheyyur Jaya Anand \textsuperscript{\rm 2},
    Chinmayi Appajigowda \textsuperscript{\rm 2}, \\
    Biplav Srivastava \textsuperscript{\rm 1},
    Lokesh Johri \textsuperscript{\rm 2},
}

\affiliations{
    \textsuperscript{\rm 1} AI Institute, University of South Carolina, Columbia, South Carolina, USA\\
    \textsuperscript{\rm 2} Tantiv4, San Jose, California, USA \\
    bharath@email.sc.edu,
    anand@tantiv4.com,
    chinmayi@tantiv4.com, biplav.s@sc.edu, lokesh@tantiv4.com
}

\maketitle

\begin{abstract}
    The state identification problem seeks to identify power usage patterns of any system, like buildings or factories, of interest. In this challenge paper, we make power usage dataset available from 8 institutions in manufacturing, education and medical institutions from the US and India, and an initial unsupervised machine learning based solution as a baseline for the community to accelerate research in this area.
    % \bips{Need to revise} Time series forecasting is an important aspect of machine learning since it has a wide range of applications, including forecasting power usage, traffic, and air quality, to name a few. For all active participants in the electrical market, load forecasting has become an important aspect of planning and operation. The penalty costs for under or over contracting electricity have increased dramatically as a result of the new market structure, making prediction error minimization more crucial than ever. Forecasting end user load for power purchase has become critical in the context of electricity suppliers. Additionally, abnormal power usage detection can be helpful in detecting excessive unwanted power usage and sometimes hinting on the device failures. 
\end{abstract}

\section{Introduction}

The growth in the deployment of Internet of Things (IoT) sensors across different industries has opened several opportunities for the economy. One of them is the collection of IoT data that companies can use to build smarter solutions. These IoT sensors, while performing their assigned tasks,  can also help collect data from real-world objects or devices for analysis and gain intelligence to improve the latter's capabilities. One such prominent application is the collection and analysis of Electricity Consumption Data (ECD) for a robust and reliable energy management system at any organization.

% Due to the growth in population and technological advancements from past few decades, there has been a increasing demand for energy requirement. With such demand in hand it is important to have a robust and reliable energy management system. 

There is a rich body of work on data-based energy management but much of it is in forecasting and with limited access to data.
% The  exists conventional methods \cite{alfares2002electric} and statistical methods \cite{tso2007predicting} to analyze and characterize the trends in the time series power data to be able to forecast the electricity consumption demand. 
Accurate forecasting of energy consumption has the potential to save large utility bills.
%great saving potential of energy utility. 
These savings can be realised once the forecasted load knowledge is used to control operations and decisions of the power utility. 
Mainly in the power systems, the economy of operations and control of operations are sensitive to forecasting errors. 
Existing forecasting methods can be characterized as conventional  \cite{alfares2002electric} or statistical  \cite{tso2007predicting} and they focus on short-term load forecasting.
% Most of the research works with statistical methods for power analysis focus on short-term load forecasting (STLF).
% The   conventional methods \cite{alfares2002electric} and statistical methods \cite{tso2007predicting} to analyze and characterize the trends in the time series power data to be able to forecast the electricity consumption demand. 
These approaches use the trends in the power data to develop a suitable model and use the model for forecasting the future load  \cite{alfares2002electric}. Based on the characteristics of the time series data, there are different statistical models popularly used. Some of them are auto-regressive (AR),  auto-regressive moving-average (ARMA), auto-regressive integrated moving-average (ARIMA). In \cite{huang2003short}, authors used Gaussian features of the load to determine the model of ARMA dynamically. 
Complex models can be used to make high-precision predictions, but this is challenging given the high complexity, irregularity, randomness, and non-linearity of real world data. Machine learning techniques can be used to create nonlinear prediction models based on a significant amount of historical data. Typical machine learning models include support vector machines (SVM) \cite{sapankevych2009time} or kernel based classification, artificial neural network (ANN) \cite{faruk2010hybrid}, tree-based ensemble methods such as gradient-boosted regression or decision trees \cite{ke2017lightgbm}, long short-term memory units (LSTM)
\cite{bedi2020energy} 
% \cite{bedi2018empirical, bedi2019deep, bedi2020energy} 
or transformers \cite{survey-timeseries-transformer}. Authors in \cite{rajapaksha2022limref} focused on providing rule based explanations for a particular forecast, considering the global forecasting model as a black-box model trained across multi-variate time series.
% Few of the works \cite{bedi2019deep,bedi2020energy} tried applying some prepossessing steps, like variance mode decomposition and optimal feature extraction using auto-encoders, to extract meaningful characteristics from the data before using a prediction model.

More recent work in energy management has focused on  
%many of the research works are focusing on
Non-Intrusive Load Monitoring (NILM) \cite{batra2014nilmtk}, \cite{hart1992nonintrusive} where, from the aggregate power data,  the aim is to 
dis-aggregate and estimate individual load. 
This technique is especially appealing to the industry due to its low cost and easy implementation.

% techniques for it's  low cost and easy implementation. 
% This technique can extract information for individual loads from total power consumption information.

Our contribution in this challenge paper for energy management are that we: (a) make a large power usage and harmonics dataset available that is collected using MiDAS sensors from Tantiv4 from 8 institutions in manufacturing, education and medical institutions from the US and India for 15 days\footnote{The duration restriction is for size reasons only. GitHub: https://github.com/ai4society/PowerIoT-State-Identification.}, and more days of data available upon request, (b) introduce and describe a generic state identification problem  that is of much interest to the industry, and (c) describe an initial un-supervised machine learning based solution to extract different operating states of the power load for a location using current harmonics that can serve as a baseline to the community to accelerate research in this area.
% \bm{needs revision} In our work, we investigated a novel ECD analysis component. We tried to extract different operating states of the power load using harmonics data for a given location. 
\section{State Identification Problem}

%   what is a state
%     subset of features with range of values

The objective of state identification (SIP) is to identify power usage patterns of any system, like buildings or factories, of interest. 
We consider it  from the perspective of power systems and as a data analysis problem. 

In an electrical system running on AC frequency, any perturbation in the system   manifests  as an energy spectrum. These spectrum patterns, or signatures, reflects the {\em state} in which the system could be: e.g., a factory floor with all machines running normally or a machine beginning to fail due to a fault, or some machines shut for repair; in a datacenter or office,
the system under heavy load; or in any industry, an unusual
load that the system has not seen till now and needs operator attention.
These signatures have traditionally been analyzed as vibration data in a technique commonly referred to as {\em Condition Based Monitoring} \cite{han2003condition} which is expensive and intrusive. 
% However, one
% could alternatively analyze the power data based on electrical
% current harmonics which offers a ’24x7’ monitoring for
% the faults or the description of the state of an electrical system.
However, one could alternatively analyze the power data based on electrical current harmonics for states of an electrical system which
offers a '24x7' monitoring for the different load patterns and faults. 
This is the true potential of the released data using Machine Learning (ML) and Artificial Intelligence (AI) methods.

As a data analysis problem, we are given a schema $F$ consisting of a list of features $F$ = \{$f_1$, .. ,$f_m$\} that a sensor is able to capture. They are also called columns in the data. 
Adopting the notations of \cite{Goodfellow-et-al-2016} for a classification problem, suppose we are also given a collection of observations $x \in \mathbb{R}^n$  corresponding to power usage at a location at a known interval $\Delta$. Each observation, or row in the data, has a structure consistent with $F$. The state identification problem is to produce a function $f$ : $\mathbb{R}^n \rightarrow$ \{$1$, ..., $k$\}. When $s$ = $f(x)$, the state corresponding to $x$ is a number capturing the category assigned to the observation. We are interested in the {\em unsupervised} SIP problem where  $k$ is unknown, i.e., the number of states has to be also learnt.

Apart from solving the SIP problem, a major challenge here is to determine metrics to use for measuring goodness. Due to the high frequency of data collection and resulting  data size, getting fully labeled ground truth is infeasible. However, users can contrast significant states and this indicates alternative ways to elicit ground truth.

%Data points can be seen as  vectors whose
%coordinates are values of variables called features.
\begin{table*}[htb]
\centering
\begin{tabular}{|l|l|l|l|l|}
\hline
\textbf{Location} & \textbf{Industry} & \textbf{Load Illustration} & \multicolumn{1}{c|}{\textbf{\begin{tabular}[c]{@{}c@{}}Size of Released\\ Dataset (15-days)\end{tabular}}} & \textbf{Load Figures (ActivePT vs datetime)*} \\ \hline
India-1 & Manufacturing &  \begin{tabular}[c]{@{}l@{}}Laser Cutting \\ Machine\end{tabular} & \begin{tabular}[]{@{}l@{}}ECD - 212MB\\ Harmonics - 3.2GB\\ 5.34\% missing data \end{tabular} &  
\begin{minipage}{0.32\textwidth}
\vspace{1mm}
\centering
\includegraphics[scale=0.32]{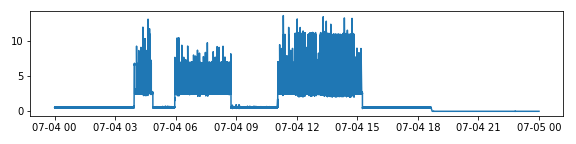}
\end{minipage}
\\ \hline
India-2 & Hospital & Main Supply & \begin{tabular}[c]{@{}l@{}}ECD - 221MB\\ Harmonics - 3.1GB\\ 1.51\% missing data \end{tabular} & 
\begin{minipage}{0.36\textwidth}
\vspace{1mm}
\centering
\includegraphics[scale=0.32]{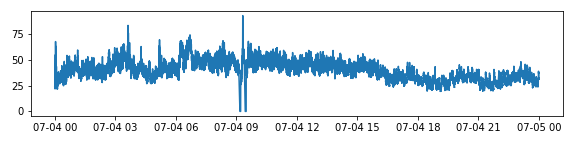}

\end{minipage} \\ \hline
India-3 & Manufacturing & Lathe Machine & \begin{tabular}[c]{@{}l@{}}ECD - 225MB\\ Harmonics - 3.3GB\\ 0.58\% missing data \end{tabular} & 
\begin{minipage}{0.36\textwidth}
\vspace{1mm}
\centering
\includegraphics[scale=0.32]{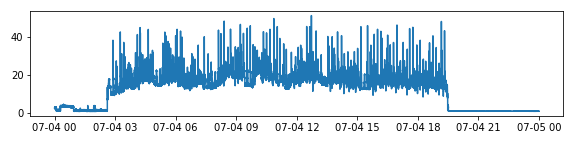}

\end{minipage}\\ \hline
India-4 & Manufacturing & Main Supply & \begin{tabular}[c]{@{}l@{}}ECD - 439MB\\ Harmonics - 3.8GB\\ 10.29\% missing data \end{tabular} & 
\begin{minipage}{0.36\textwidth}
\vspace{1mm}
\centering
\includegraphics[scale=0.32]{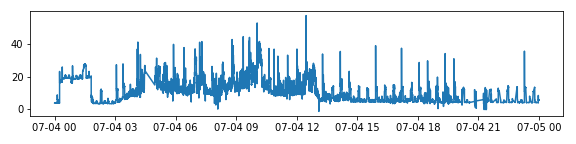}

\end{minipage}\\ \hline
India-5 & Manufacturing & CNC Machine & \begin{tabular}[c]{@{}l@{}}ECD - 266MB\\ Harmonics - 3.8GB\\ 0.13\% missing data \end{tabular} & 
\begin{minipage}{0.36\textwidth}
\vspace{1mm}
\centering
\includegraphics[scale=0.32]{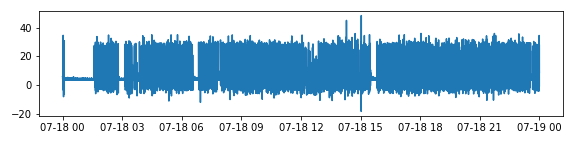}

\end{minipage}\\ \hline
India-6 & Manufacturing & Main Supply & \begin{tabular}[c]{@{}l@{}}ECD - 175MB\\ Harmonics - 2.3GB\\ 0.41\% missing data \end{tabular} &
\begin{minipage}{0.36\textwidth}
\vspace{1mm}
\centering
\includegraphics[scale=0.32]{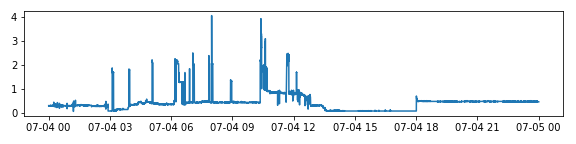}

\end{minipage}\\ \hline
USA-1 & Education & AI/ML Lab & \begin{tabular}[c]{@{}l@{}}ECD - 393MB\\ Harmonics - 3.4GB\\ 1.67\% missing data \end{tabular} & 
\begin{minipage}{0.36\textwidth}
\vspace{1mm}
\centering
\includegraphics[scale=0.32]{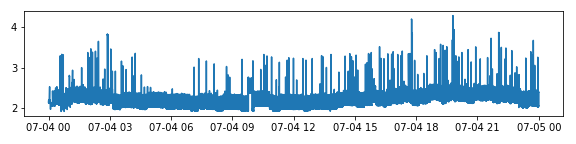}

\end{minipage}\\ \hline
USA-2 & Education & Data center & \begin{tabular}[c]{@{}l@{}}ECD - 467MB\\ Harmonics - 3.9GB\\ 1.63\% missing data \end{tabular} & 
\begin{minipage}{0.36\textwidth}
\vspace{1mm}
\centering
\includegraphics[scale=0.32]{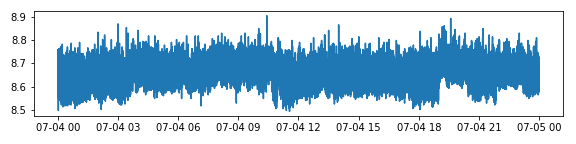}

\end{minipage}\\ \hline
\end{tabular}
\caption{Characteristics of data collected locations; *illustrative power usage for a day - Jul 04 (except India-5: Jul 18, 2022).}
\label{tab:location-comparision}
\end{table*}

\section{Data Collection and Usage}

% In the current project, we have used data that is 
We now discuss how the data is collected by the MiDAS device installed by Tantiv4.

\subsubsection*{MiDAS IoT Sensors:} 
%\bips{Chinmayi, Anand, Lokesh - please add}.

% details about sensor description, differencing aspect of the sensor when compared with others in the market, 

The MiDAS IOT sensor measures phase voltages (three-phase), phase currents (three-phase), neutral current, power factor (three-phase), active power (three-phase), apparent power (three-phase), reactive power (three-phase), frequency and phase (three-phase) values every 300ms. The device is also capable of collecting three-phases of current and voltage harmonics data from 2 to 32 harmonic levels along with total harmonic distortion for each phase of current and voltage every 500ms. It interfaces using current sensors with a clamp format for easy installation. Voltage sensors are internal to the device. Field terminals can take up to 1.5 sq. mm cables. 

\subsubsection*{Data Collection:} 
To demonstrate the generality of the SIP problem, we are making datasets available following the FAIR guiding principles \cite{wilkinson2016fair} from different economic industries: Manufacturing, Hospital \& Educational institutions. The instructions to obtain the dataset are available here (https://github.com/ai4society/PowerIoT-State-Identification) along with the documentation and the metadata for the released dataset. The GitHub documentation contains a google form link, where the user can fill in the google form and the dataset link can be obtained. The need for the form is that we can inform the user about any data update. The dataset obtained from MiDAS device is of two forms: Electricity Consumption Data and Harmonics Data. The harmonics information obtained by the MiDAS sensor can be used for further analysis of electric device performance.

\textbf{Electricity Consumption Dataset:} contains 28 different features of electricity consumption data. The features are: Current (IA, IB, IC INCURRENT), Voltage (VA, VB, VC), Power Factor (PFA, PFB, PFC, PFT), Phase (PhaseA, PhaseB, PhaseC), Active Power (ActivePA, ActivePB, ActivePC, ActivePT), Reactive Power (ReactivePA, ReactivePB, ReactivePC, ReactivePT), Apparent Power (ApparentPA, ApparentPB, ApparentPC, ApparentPT), Frequency (FREQ), and Time Stamp.

\textbf{Harmonics Dataset:} contains 193 different harmonics data features. The features are: Current (AI\_HR[2 to 32], AI\_THD, BI\_HR[2 to 32], BI\_THD, CI\_HR[2 to 32], CI\_THD), Voltage (AV\_HR[2 to 32], AV\_THD, BV\_HR[2 to 32], BV\_THD, CV\_HR[2 to 32], CV\_THD) and Time Stamp.

The characteristics and the types of different loads present at different locations is shown in Table \ref{tab:location-comparision}. Active power characteristics plotted for a single weekday for different locations is shown in 'Load Figures' column in Table \ref{tab:location-comparision}. Electricity consumption and harmonics data-sets for all of the locations listed in Table \ref{tab:location-comparision} are made available online for 15 days for the period July 01, 2022 to July 15, 2022 (except for location India-5: July 16, 2022 to July 31, 2022). The dataset sizes for each location for 15 days are shown in 'Size of the Released Dataset' column of Table \ref{tab:location-comparision}. Beyond what is already being released, we have data available for these locations from January-August 2022. So, additional data for more days can be obtained for research purposes by contacting the authors.

\begin{itemize}
\item \textbf{India-1:} 
%The  sensor is connected to Laser Cutting Machine. 
The sensor is connected to a single Laser cutting machine with  power readings between 2 to 25 Amps. The machine laser cuts stainless, carbon steel, aluminum, brass, titanium, and more.

\item \textbf{India-2:} The sensor is connected to a hospital's  main incoming supply and the power load includes  machines for CT scan, ECG, EEG, Digital Xray, USG, and C-ARM diagnostic services. The power consumption fluctuates between 35 to 110 Amps. 

\item \textbf{India-3:} The sensor is connected to a single device lathe machine which is used to perform various operations such as cutting, sanding, knurling, drilling, deformation, facing, and turning, with tools that are applied to the work piece to create an object with symmetry about that axis with power consumption between 2 to 40 Amps.

\item \textbf{India-4:} The sensor is connected to the main supply of a manufacturing plant which comprises of devices such as multiple CNC (computer numerically controlled) machines, Lathe machines, Lifts etc. and the power consumption is between 15 to 60 Amps.

\item \textbf{India-5:} The sensor is connected to a single CNC machine which is used for testing roughness, waviness, flatness, curvature etc of objects and the power consumption is between 3 to 25 Amps.

\item \textbf{India-6:} The sensor is connected to the main supply of design \& drafting division, comprising of less than 10 employees, equipped with dedicated plotters, jumbo photo copiers, blue printer, spiral binder etc. and the power consumption is between  0.5 to 10 Amps.

\item \textbf{USA-1:} The sensor is connected to the main supply of a research center at a University with  10-30 daily users who bring their devices or use servers.

\item \textbf{USA-2:}  The sensor is connected to the main supply of a server room at a Computer Science department of a University being used for various computational loads.

\end{itemize}

\subsubsection*{Data Cleaning and Augmentation:}

While working with sensors in the real world, there are often times when data is missing. 
% Many real world sensors face with the problem of missing data. 
%There can be numerous reasons for missing data. 
This can be due to various reasons such as power outage, sensor failure, sensor maintenance work or the sensor is not connected to the network. If one ignores missing data, the quality of analysis, as measured by metrics like accuracy, with them could be degraded.
Hence, we employ the optional step of processing it (detailed in code documentation). A user could skip / substitute it with their own methods for cleaning and augmenting the data.
In the released data, we filled in a timestamp's missing data by taking the mean of two previously and subsequently available occurrences of the same timestamp. For example, if the data is missing at 02-01-2022 09:00:00, we looped back(forward) to the previous(future available) dates where the data is available at 09:00:00 timestamp and considered mean of these observations. The main idea of considering the same timestamp is that, given a working environment, the characteristics of the power consumption data for a given timestamp would be similar across the working/non-working days. 
\section{Initial Solution - Power State Clustering}

\begin{figure}[htbp]
    \centering
    \includegraphics[scale=0.30]{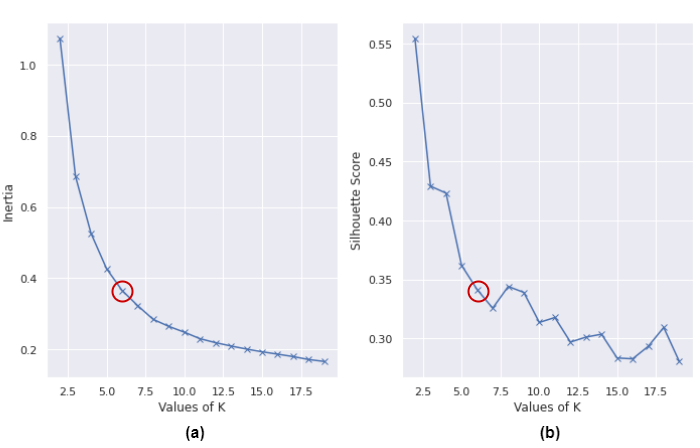}
    \captionsetup{justification=centering}
    \caption{For location India-4 (a) Elbow curve plot with inertia values for different number of clusters (b) Silhouette score plot shows that the optimal number of clusters are 6.}
    \label{fig:elbow_sil}
\end{figure}

With the help of harmonics data that is being collected by MiDAS sensor, system load performance can be analyzed in depth. This provides more insights regarding the power usage especially in an industrial setting. 
%The following section will 
We will now discuss the power state clustering method that is used to cluster the power usage data into  power states using the harmonics data, with the clusters signifying the system states.

\subsection{Data Used}
The harmonics data collected by the MiDAS sensor consists of 32 harmonic values for three-phases of current and voltage. In our experiment, we have considered the odd harmonics of current for all the three-phases. This feature dataset is sampled at 1 minute frequency. We will use data from India-4 for illustration. 

\subsection{Finding Number of States For a Location}

To classify the different harmonic observations into state categories of system, we must first determine what are the different states that are present in a system for a given time interval. These different system states can be discovered by applying a  {\em clustering algorithm} over the selected feature dataset. 

\subsubsection{Clustering Method:}
%In various domains, such as machine learning, data mining, pattern recognition, image analysis, and bio-informatics, clustering is a widely used statistical data analysis method.  
Clustering is a widely used statistical data analysis method  in which similar items are grouped into various groups, or more specifically, the data to be analyzed is partitioned into subsets so that the data in each subset are based on a predetermined distance metric \cite{madhulatha2012overview}. We experimented with multiple clustering methods (K-Means, DBScan and Agglomerative) and only present the best result obtained with the partition based K-Means.
% The clustering algorithm presented  for our experiments fall is partition based (K-Means clustering). 
% \subsubsection*{Agglomerative Clustering} is a method of hierarchical based clustering. This is also called bottom-up clustering as it starts by considering each data point as an individual cluster. As we advance, the distance between the each cluster is calculated and the clusters with shortest distance are merged. The process continues until the desired number of clusters is reached. 

% The selection of a distance metric is a crucial stage in a hierarchical clustering. For our experiments, we have used the euclidean distance metric. The formula for calculating euclidean distance between two points ($x_1$,$y_1$) and ($x_2$,$y_2$) is shown in equation \ref{eq:dist}

% \begin{equation}
% d = \sqrt{ (x_1 - x_2)^2 + (y_1 - y_2)^2 } 
% \label{eq:dist}
% \end{equation}

\subsubsection*{K-Means Clustering:} is a partitioning based or centroid-based clustering method. 
%In contrast to hierarchical clustering, this form of clustering arranges the data into non-hierarchical clusters. 
Partitioning algorithms work by defining an initial number of groups and then iteratively reallocating items among them to achieve convergence. Typically, this technique selects all clusters simultaneously \cite{madhulatha2012overview}. 

\subsubsection*{Determination of optimal number of clusters}
Prior to running the clustering algorithm, many clustering techniques require the specification of the number of clusters to be produced in the input data set. Many techniques have been proposed for this problem in literature - one such method is using the elbow curve method. 
% Alongside with elbow curve method we have also inspected the clusters formed using silhouette score metric\biplav{Give citation}.
Besides this, we also inspect the quality of clusters using the silhouette score \cite{shahapure2020cluster}.

\subsubsection{Elbow Method:}
According to the elbow criterion, one should select the number of clusters such that including another cluster would not provide significantly more additional information \cite{madhulatha2012overview}. 
To chose the optimal number of clusters for a given dataset, we have used the elbow curve method. The sum of the squared distances between the samples and the cluster center is known as {\em inertia}. Lower values of inertia indicate that the clusters are well separated. But the inertia becomes 0 once the number of clusters is equal to number of samples. Considering this trade-off between the inertia and the number of clusters, we manually chose the elbow point from the inertia graph. With varying number of clusters from $n=1$ to $n=20$, K-Means algorithm is performed on a 1 week data (January 10, 2022 to January 15, 2022) for each $n$ value and the respective inertia value is calculated. This inertia graph for the respective number of clusters is shown in Figure \ref{fig:elbow_sil}(a). From Figure \ref{fig:elbow_sil}(a), it can be observed that after this elbow point, there is no much change in the inertia value for increasing number of clusters. 
Hence the elbow point is chosen as the optimal number of clusters. Here the value of optimal number of clusters is falls between 6 to 8. 

\begin{figure*}[!t]
    \centering
    \captionsetup{justification=centering}
    \includegraphics[scale=0.55]{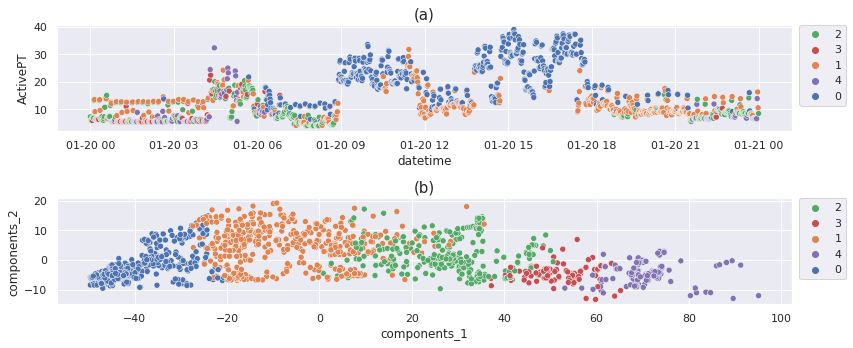}
    \caption{Classifier output for location India-4, January 20, 2022 (Thursday) (a) Clusters (states) viewed with Active Power on Y-axis
    (b) PCA for 2-dimensional visualization of the clusters (states).}
    \label{fig:clusters}
\end{figure*}

\begin{figure*}[!t]
    \centering
    \captionsetup{justification=centering}
    \includegraphics[scale=0.55]{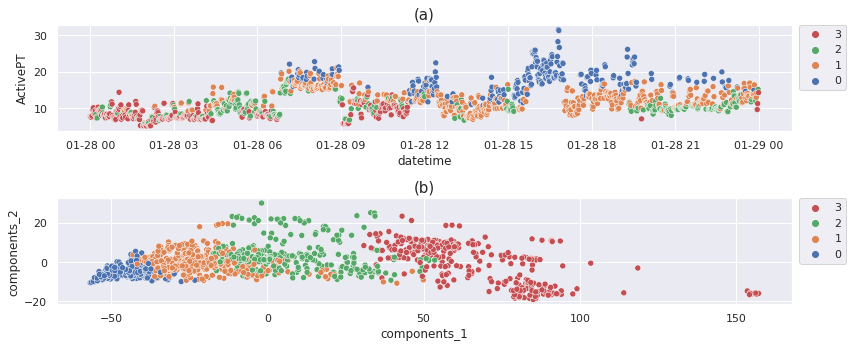}
    \caption{Classifier output for location India-4, January 28, 2022 (Friday) (a) Clusters (states) viewed with Active Power on Y-axis
    (b) PCA for 2-dimensional visualization of the clusters (states).}
    \label{fig:clusters_new}
\end{figure*}

\subsubsection{Silhouette Score:}
The silhouette coefficient is calculated for each data point by
$s$ = $\frac{(b - a)}{\max(a,b)}$, where
% using Equation \ref{eq:sil}.
\textit{a} is the mean intra-cluster distance and \textit{b} is the mean nearest-cluster distance \cite{shahapure2020cluster}. Silhouette score is the mean of silhouette coefficient calculated for all the samples. This score ranges from +1 to -1, and the higher the score, the better the clustering.
% \begin{equation}
% s = \frac{(b - a)}{\max(a,b)}
% \label{eq:sil}
% \end{equation}
According to the silhouette score analysis and the elbow curve method, it is found that the optimal number of clusters is 6. 

\subsubsection{Dimentionality Reduction}
As we have discussed previously, the data that is being used for clustering consists of 15 features (considering only odd harmonics of current). Visualizing the clusters formed using this high dimensional data is hard to interpret. Therefore, we have used one of the popular dimentionality reduction algorithm, Principle Component Analysis (PCA) \cite{song2010feature}. We have extracted two principle components using the PCA algorithm for a 2-dimensional visualisation of the clusters formed. 
\\
\\

\vspace{-0.2in}
\subsection{Assigning States to Power Data of a Location}

Once the different system states are identified for a given time interval, we can build a classification model to classify the collection of observations $x$ into their respective state categories $f(x)$. For this experiment we have used the random forest classifier, which is a tree-based classification algorithm \cite{breiman2001random}. The training dataset comprises of 3 weeks of January (Jan 03, 2022 to Jan 23, 2022). The system state categories are identified %for each individual week 
using the clustering algorithm and these state categories along with the feature dataset can be used to train a classifier to predict states. We use the random forest classifier for the latter. 

% After training the model with the system state categories data, we have tested the performance of the model in classifying the different system states. 

We now illustrate the usage of the  trained model.
In Figure \ref{fig:clusters} and Figure \ref{fig:clusters_new}, we show the classification results which correspond to the identified states at the location for two different dates. We note that 5 states were identified in the former for Jan 20 (Thursday) while 4 states are identified in the latter for Jan 28 (Friday). The order of the legend corresponds to the descending order of state labels based on the number of data points clustered into the respective state label. The performance of the classification model is evaluated using F1 score metric. The F1 scores of the system state predictions for the days Jan 20 (Thursday) and Jan 28 (Friday), for which the system labels have been reported in Figure \ref{fig:clusters} and Figure \ref{fig:clusters_new}, are 0.51 and 0.67 respectively.
% The first section of the each Figure represents the different states classified over active power on y-axis. The second 

% The system states obtained after classification for Jan 20 and Jan 28 are shown in Figure \ref{fig:clusters}(a) and Figure \ref{fig:clusters_new}(a) respectively. 

% For illustration, we have chosen a two different days, Jan 20, 2022 (Thursday) and Jan 28, 2022 (Friday), from location India-4 as our test data-sets. The trained model is used to forecast the system state categories for the selected test dataset. The system states obtained after classification for Jan 20 and Jan 28 are shown in Figure \ref{fig:clusters}(a) and Figure \ref{fig:clusters_new}(a) respectively.  

\section{Discussion and Challenges}

In this paper, we presented a rich dataset of power consumption spanning 8 locations in 3 industries (manufacturing, education and hospital) and two countries (India, US) for 15 days  whose size exceeds 27GBs. 
In addition, more days of data for the same locations from January-August 2022 can be requested for research purposes by contacting the authors.

This data can be used for many energy management applications. While the most studied use-case in literature is energy forecasting for which this data can also be used, we envisage its significant potential while tackling  the  introduced  state identification problem - our second contribution. The SIP problem is driven by industry need.
%that can be tackled with the released data.

Furthermore, we described a baseline approach for SIP and presented results for India-4.
% of the experiments performed on location India-4. 
For this, we performed K-Means clustering by providing the optimal number of clusters as k=6 and the clustering performance 
%of this algorithm 
is evaluated using the silhouette score metric. These clusters formed can be directly  correlated with the system states $f(x)$ for a given collection of observations $x$. Figures \ref{fig:clusters} and \ref{fig:clusters_new} show the different system states obtained from the classification model, trained over the system state labels obtained from the K-Means algorithm, for two different days. Each cluster label in  Figures \ref{fig:clusters} and \ref{fig:clusters_new} represents the category to which the system state belongs to. Due to the higher dimensionality of the dataset, we have used PCA to extract two principle components (component\_1 and component\_2) from the feature dataset. In Figure \ref{fig:clusters}(b), these two principle components are plotted on x-axis and y-axis, respectively, along with the labels obtained for different states from the classification model. We can observe that the different system states are grouped distinctly across the principle component space. The results were validated with domain experts on the ground.
% We can observe that the different system states are grouped distinctly across the principle component space. The results were validated with domain experts on the ground.

Researchers using our dataset to identify different states of the systems (locations)  with their own methods can validate their results by following the documentation provided in the \textit{data} folder of our GitHub repository and plots in the {\em results} folder. This folder contains CSV files with different states identified  on each day by location along with their centroid details . The best F1 score  of the classification model, by dates, can be found in the \textit{leaderboard} folder.

Our work opens up many research challenges to the community. First, one needs to handle large data sizes containing time-series and harmonics information, and tackle missing data. Second, one needs to create effective algorithms to identify states in an unsupervised or minimal labeling settings. Third, the methods have to be general across applications and we provided data from three different industries. Fourth, the results should be understandable to the user for which innovations are needed in visualization, interaction and explanation generation. We have begun initial steps towards them  \cite{midas-forecasting,nl2sql}.  

% More generally, the many categories of the system states can be related with the clusters generated using the current harmonics. But the key problem would be to explain the nature of these state categories obtained for a system. Explaining these categories would be helpful for understanding the power consumption pattern of a system.

%\input{content/conclusion}

\bibliography{references}

\end{document}